\documentclass[english,aip,jcp,reprint]{revtex4-2}

\usepackage[T1]{fontenc}
\usepackage[utf8]{inputenc}
\usepackage{babel}
\usepackage{amstext}
\usepackage{graphicx}
\usepackage[unicode=true,pdfusetitle,
 bookmarks=true,bookmarksnumbered=false,bookmarksopen=false,
 breaklinks=false,pdfborder={0 0 1},backref=false,colorlinks=false]
 {hyperref}

\makeatletter
\usepackage[version=4]{mhchem}
\usepackage{notes2bib}

\bibnotesetup{
note-name = ,
use-sort-key = false
}

\usepackage{cleveref}
\AtBeginDocument{%
\let\ref\cref
}

\crefname{enumi}{observation}{observations}

\usepackage{textcomp}

\@ifundefined{showcaptionsetup}{}{%
 \PassOptionsToPackage{caption=false}{subfig}}
\usepackage{subfig}
\makeatother

\begin{document}
\title{A call to arms: making the case for more reusable libraries}
\author{Susi Lehtola}
\email{susi.lehtola@alumni.helsinki.fi}

\affiliation{Department of Chemistry, University of Helsinki, P.O. Box 55, FI-00014
University of Helsinki, Finland}
\begin{abstract}
The traditional foundation of science lies on the cornerstones of
theory and experiment. Theory is used to explain experiment, which
in turn guides the development of theory. Since the advent of computers
and the development of computational algorithms, computation has risen
as the third cornerstone of science, joining theory and experiment
on an equal footing. Computation has become an essential part of modern
science, amending experiment by enabling accurate comparison of complicated
theories to sophisticated experiments, as well as guiding by triage
both the design and targets of experiments and the development of
novel theories and computational methods. 

Like experiment, computation relies on continued investment in infrastructure:
it requires both hardware (the physical computer on which the calculation
is run) as well as software (the source code of the programs that
performs the wanted simulations). In this Perspective, I discuss present-day
challenges on the software side in computational chemistry, which
arise from the fast-paced development of algorithms, programming models,
as well as hardware. I argue that many of these challenges could be
solved with reusable open source libraries, which are a public good,
enhance the reproducibility of science, and accelerate the development
and availability of state-of-the-art methods and improved software.
\end{abstract}
\maketitle
\newcommand*\ie{{\em i.e.}}
\newcommand*\eg{{\em e.g.}}
\newcommand*\etal{{\em et al.}}
\newcommand*\citeref[1]{ref. \citenum{#1}}
\newcommand*\citerefs[1]{refs. \citenum{#1}} 
\newcommand*\Citeref[1]{Ref. \citenum{#1}}

\newcommand*\Erkale{{\sc Erkale}}
\newcommand*\Bagel{{\sc Bagel}}
\newcommand*\FHIaims{{\sc FHI-aims}}
\newcommand*\HelFEM{{\sc HelFEM}}
\newcommand*\Gaussian{{\sc Gaussian}}
\newcommand*\LibXC{{\sc LibXC}}
\newcommand*\Orca{{\sc Orca}}
\newcommand*\PySCF{{\sc PySCF}}
\newcommand*\PsiFour{{\sc Psi4}}
\newcommand*\Turbomole{{\sc Turbomole}}

\section{Introduction \label{sec:Introduction}}

All software is aging software, and all code becomes legacy code at
some point. No matter how good source code you write now, it will
become outdated in a matter of years to decades because of the following
three fundamental observations:
\begin{enumerate}
\item Improved scientific models and mathematical algorithms are continuously
developed. \label{enu:new-and-more}
\item Computer hardware is constantly evolving. \label{enu:computer-hardware-is}
\item Computer programming models and programming languages likewise keep
evolving. \label{enu:computer-programming-models}
\end{enumerate}
On all three counts, scientific software is therefore trying to hit
a constantly moving target, and staying up to date requires constant
maintenance and development.

Related to \ref{enu:new-and-more}, a present-day state-of-the-art
implementation becomes obsolete, when an improved (faster or more
accurate) algorithm is discovered. The choice of the algorithm is
arguably the most important piece in the design of scientific software,
and a significant part of the advances made in computational science
are foremostly thanks to access to improved algorithms rather than
simply to the availability of more computational power. The choice
of the algorithm determines the asymptotic scaling of the calculation,
and an unoptimized implementation of the best algorithm is oftentimes
much faster than a heavily optimized implementation of a subpar algorithm.
This is why keeping track with new algorithms and implementing them
in software is perhaps the most important consideration relating to
scientific software, and this requires constant effort.

Related to \ref{enu:computer-hardware-is}, a central challenge in
scientific computing is that present-day (super)computers look nothing
like they used to. For a long time, computers were built around a
single processor core, which became faster and faster every few years
thanks to Moore's law.\citep{Moore1998_PI_82} Because the clock speed
also kept going up, this meant that over a time span ranging several
decades, old programs ran faster and faster on newer and newer hardware,
without ever having to make any changes to their source code. In this
era, most development efforts were focused at making codes run as
fast as possible on the single processor core. 

An important observation related to these two developments was recently
made by \citet{Lehtola2022_WIRCMS_1610}: commodity hardware in the
present day is as fast as the fastest supercomputer in the world in
the 1990s. Combined with the present-day availability of faster algorithms,
then-pioneering calculations can nowadays be reproduced even on students'
laptop computers with easily installable free and open source software
(see \citeref{Lehtola2022_WIRCMS_1610} for the employed definition
and more details).

However, because higher clock speeds inherently mean higher power
consumption, already many years ago there was a paradigm shift from
single-core to many-core architectures. Instead of making the single
core run faster, it became more power-efficient to add computing power
by introducing more processor cores. This development already introduced
changes to programming models. Old codes no longer necessarily ran
faster on a new computer than on an old one, since the speed of a
single core stagnated or even decreased. Instead, using the power
of a new machine suddenly required leveraging all of its processor
cores, thus requiring the use of parallel programming paradigms, such
as open multi-processing (OpenMP) shared-memory parallellism.

But, the evolution of hardware did not stop there. Modern high-performance
computers employ non-uniform memory architectures (NUMAs), which cannot
be efficiently targeted with shared-memory parallellism; distributed-memory
programming models such as the message passing interface (MPI) are
instead required, accompanied with the use of distributed computing
algorithms which are considerably more difficult to develop than their
sequential or shared-memory counterparts.

A further development is the advent of general-purpose graphics processing
units (GPGPUs), which have become a quintessential part of the fastest
available supercomputers. For instance, 7 out of the 10 fastest systems
on the most recent top-500 list of supercomputers\bibnote{\url{https://www.top500.org/lists/top500/2023/06/}, accessed 5 Oct 2023.}
rely on GPGPUs, and only the Fugaku\citep{Sato2022_IM_26} (\#2 on
the list), Sunway Taihulight\citep{Fu2016_SCIS_72001} (\#7 on the
list), and Tianhe-2A\citep{Liao2014_FoCS_345} (\#10 on the list)
supercomputers do not use GPGPUs. Still, achieving good performance
on even these three non-GPGPU systems requires employing implementations
reminiscent of those optimal for GPGPU systems, as well. 

As a result, good GPGPU support has become an important consideration
in scientific software development. Unfortunately, classical algorithms
designed for CPUs often run poorly on GPGPUs, and the algorithms usually
need to be redesigned into forms that are more vectorizable, have
higher arithmetic intensity, and are more suitable for massively parallel
computation. As an in-depth discussion on this topic is beyond the
scope of this work, I refer the reader to the recent perspective by
\citet{Felice2023_JCTC_} and references therein for further analysis.

In summary, taking advantage of the full power of even a single node
of a modern supercomputer therefore requires efficient utilization
of both NUMA and GPGPU aspects, which have not been traditionally
targeted by scientific software developers, as these platforms have
only recently become important. Indeed, many scientists still privately
admit hoping that they could forget altogether about the intricacies
of NUMA and GPGPUs, and instead to continue targetting traditional,
simpler-to-program shared-memory computers. In the present day, such
a strategy unfortunately limits scalability to low-end commodity hardware.

The development of programming languages, \ref{enu:computer-programming-models},
has partly addressed the aforementioned issues. For instance, the
OpenMP programming model greatly simplifies code development for shared-memory
architectures: in the extreme case, legacy software can be efficiently
parallellized with OpenMP by the addition of a few critically placed
parallel loop statements. 

The introduction of new standards and libraries for old programming
languages typically allows for more expressive code: a given task
can often be accomplished in newer language standards with considerably
fewer lines of source code, which is better for code readability and
developer productivity. This can also be of huge assistance in keeping
the code up to track with new computational methods and algorithms:
when the code is faster to develop, it is also easier to keep up to
date in methodology.

New programming languages are also introduced to simplify the development
and maintenance of new software. For instance, Rust and Julia have
recently gained use in scientific software.\citep{Perkel2019_N_141,Perkel2020_N_185}
The former aims for an inherently safe programming paradigm by eliminating
the potential of memory errors while ensuring high performance, while
the latter simplifies the support of new types of computing hardware
by automatization: Julia code can be automatically run on GPGPUs.

However, like in the human world, also in the computer world there
is a language barrier: it is not always straightforward to get programs
and libraries written in different languages to talk with each other.
The difficulty of interfacing to state-of-the-art libraries written
in another language often leads to the need to reimplement the missing
functionality in the same language as your program, thus limiting
reusability of code across languages. Design patterns and program
structures, however, may be reusable.

The golden standard of interfaciability is the C language. Any problem
that is straightforward enough to be expressible in terms of a C interface
can be approached with a combination of languages, as most languages
are interoperable with C (e.g. Rust, Julia, CPython, and Fortran 2003
with the \texttt{iso\_c\_binding} module). 

In contrast, there is no general solution to interface object oriented
libraries written in different languages. This problem has only been
solved for some special combinations: for instance, C++ code can be
made accessible from Python, but this requires developing specialized
wrappers in the C++ code with packages like pybind11\citep{Jakob2017__}
or cppyy\citep{Lavrijsen2016__}. Unfortunately, the reverse---accessing
Python from compiled languages such as C, C++ or Fortran---is rarely
done in practice as it is considered awkward, delicate, and difficult
to debug. Although the Python language is extremely useful in many
applications, this lack of access to Python functionality poses severe
limitations on the reusability of Python code in other languages.

The interfaciability of Rust and Julia, in turn, appears to be presently
limited to C bindings. Although the situation is likely to change
in the upcoming decade, the reusability of Rust and Julia programs
therefore remains likewise currently limited.

The C++ language is the \emph{de facto} standard for new high-performance
computing projects, with the venerated Fortran language---which is
the overwhelmingly dominating language in legacy codes---still holding
second place. Good code can be written in any language; some languages
just are designed to better allow the development of clean and maintainable
code. C++ and modern Fortran are in principle both good options for
developing scientific software in the present day.

Another important issue to consider is software stack support on supercomputers.
As the lifetime of supercomputers is several years, compiler support
for new programming paradigms is often not straightforwardly available
on older installations. Likewise, support for new languages like Rust
or Julia is often limited on older systems and may require the installation
of a large software stack, which tends to be unappealing to systems
administrators, and demands expertise from the end user.

While the software management challenge has been recently addressed
via the introduction of cross-platform package managers such as conda,\citep{Anaconda__}
a recurring issue with such third-party packages is that the end user
has no guarantee that the binaries installed by the package manager
are optimal for the used platform. Although slightly suboptimal binaries
are not a huge issue in amateur use as discussed by \citet{Lehtola2022_WIRCMS_1610},
much more consideration is often necessary in high-performance environments.
When a program is going to be run for millions of CPU hours, it becomes
really desirable to employ optimal binaries, which typically means
that the end user has to be able to compile the software from source
code. 

A strategy often used in scientific software development is to conciously
hold back to older standards (such as C++11 instead of the newest
C++20 standard, or Fortran 2003 instead of the newest Fortran 2018
standard) that are natively supported by the system compilers on various
companies and universities' computer clusters and supercomputers,
as this can significantly ease the human-side considerations of software
maintenance: such software is easy to compile on any currently system
that is currently running. The choice of the language standard can
therefore always be seen as a compromise between the access to the
power and features of newer standards \emph{vs}. the arising requirements
for recent compilers and libraries that may severely limit the attractiveness
of a package for codes targetting legacy installations.

\begin{figure}
\subfloat[Initial idea for research code. Only the starting point and the final
goal are known. The steps necessary to accomplish the goal or the
layout of the program are still unknown. \label{fig:Initial-plan-for}]{\begin{centering}
\includegraphics[width=0.6\linewidth]{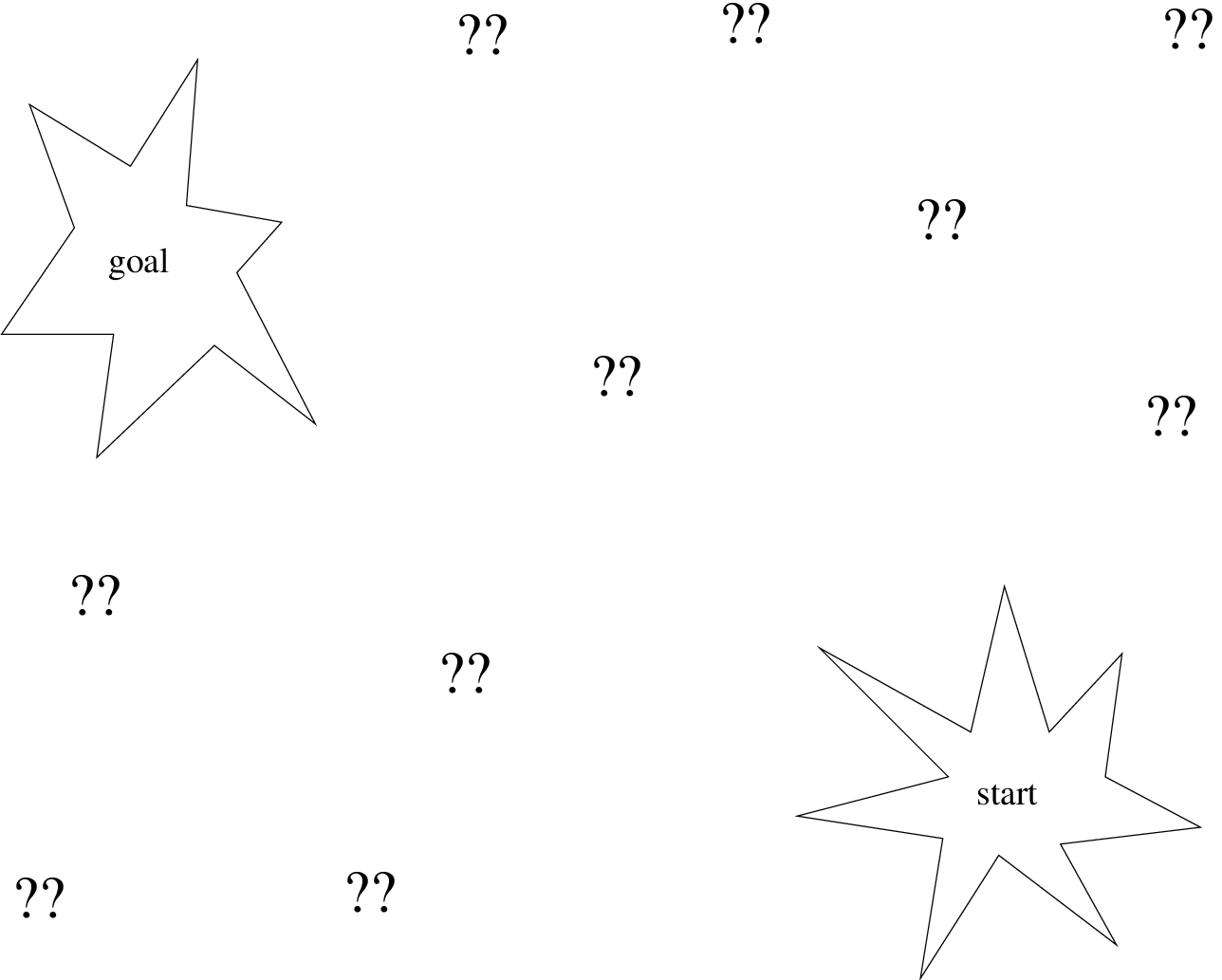}
\par\end{centering}
}

\subfloat[First version of research code. During the development of the code,
the researcher finds out by trial and error what steps are necessary
(pathfinding and trailblazing). The numbering is out of sequence to
illustrate that the design is often iterative, and can require revisiting
earlier steps of the algorithm. As a result of the organic development,
the structure of the code is not clear. \label{fig:First-version-of}]{\begin{centering}
\includegraphics[width=0.6\linewidth]{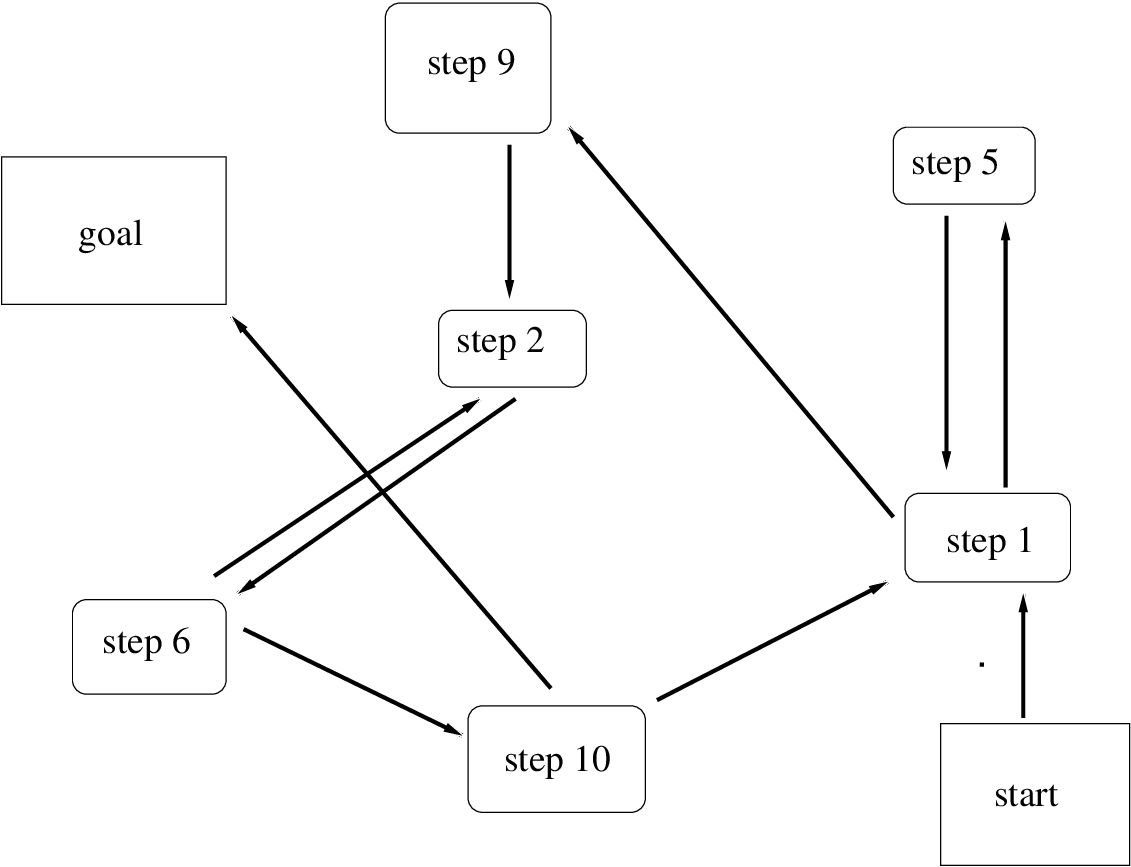}
\par\end{centering}
}

\subfloat[Production-grade code. As the result of extensive refactoring and
a deeper understanding into the task to be solved, the researcher
has found the simplest and cleanest approach to arrive at the wanted
goal. The solution has been divided into modules that accomplish clearly
defined tasks. \label{fig:Production-grade-code}]{\begin{centering}
\includegraphics[width=0.6\linewidth]{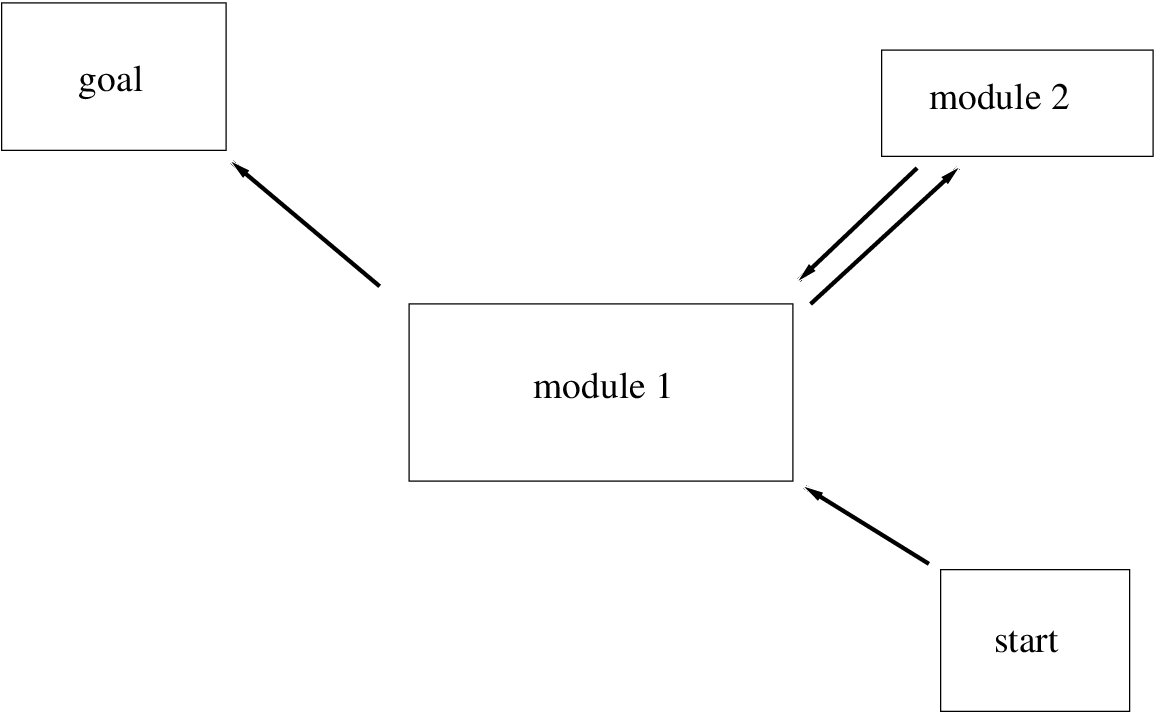}
\par\end{centering}
}

\caption{Illustration of a typical scientific coding project. \ref{fig:Initial-plan-for}:
the developer starts with only a vague idea of what the code should
accomplish. \ref{fig:First-version-of}: the first version is functional,
but may be overly complicated. \ref{fig:Production-grade-code}: after
many refactors, the production grade code offers the cleanest possible
structure. \label{fig:Illustration-of-a}}
\end{figure}

Assuming that a language standard such as C++11 or Fortran 2003 has
been chosen, what does typical scientific code development look like?
This is illustrated in \ref{fig:Illustration-of-a}, and will be discussed
in detail in the next section.

\section{Scientific Code Development \label{sec:Code-Development}}

\subsection{Research Code vs Production Grade Code \label{subsec:production}}

When talking about software in science, it is essential to differentiate
between research code and production grade code; yet, this distinction
is rarely made in the literature. Most code developed in academia
is research code: code that is being developed to solve an immediate
research problem. 

It should furthermore be pointed out here that the overwhelming majority
of research code is developed for a single use, and then discarded
afterwards. An essential part of computational science is the need
to analyze the output of simulations. Bash, Python, and Matlab scripts
are commonly used for this purpose. Scripts are written not only by
professional scientific software developers, but also the larger research
community---who are far more numerous than professional software
developers---to analyze the results obtained with others' software.
Scripts therefore likely represent the overwhelming majority of all
research codes. As the requirements in each project are different,
these scripts are not used beyond a single project.

Let us now examine the case of scientific code development to solve
a more complex scientific problem. In such a case, at the beginning
one usually does not even know what the code should look like (\ref{fig:Initial-plan-for}).
Instead, one just has to start from somewhere, and find their way
while writing the code. Although the thought of being able to design
the perfect program on a piece of paper before writing a single line
of code is beautiful, it is also a utopia and can easily lead one
to procrastination and not accomplishing anything.

In reality, one has to bite the bullet and employ an iterative approach
to code development. The first try is like finding your way in the
dark: it may not be pretty, but if you keep at it, it will eventually
get you where you want to go (\ref{fig:First-version-of}). Often,
the hardest step is just to get started. Powerful scientific programming
languages such as Python lower the height of this first step---trailblazing
and pathfinding---by enabling fast prototyping. A high-level language
with plentiful libraries for a variety of purposes makes it easier
to write an initial implementation and figure out what exactly should
be done, and how.

Research code is never complete: new types of methods and ways to
do things are always being explored, requiring novel code to be developed
and old code to be modified. This means that research code tries to
reach goals that are ever evolving. If you develop new methods or
algorithms, you are in uncharted territory and are again in the dark.
And once that task has been completed (the paper has been published),
the code is likely discarded, or archived with the rest of the project's
data.

There will always be new research codes, because the goal is moving.
Production grade code is different. Provided that the end goal found
in a research-stage code is either important by itself or useful in
another context as well, the next stage is to streamline the path:
what is the shortest path from the start (input) to the end (output)?

Production grade code is therefore written to solve a specific, well-defined
task, and typically arises through years of experimentation by various
authors in many research codes.\citep{Adorf2019_CSE_66} Although
a new research project begins in the dark---the software goals are
unclear at the beginning---by the point where writing production
grade code becomes relevant, many people have already experimented
with different types of algorithms and written various implementations.
Once many paths have been explored, a consensus has arisen on (i)
what is the task to be accomplished, and (ii) what are good ways to
achieve this task. Production grade code therefore has a clearly defined
mission: solving some important (sub)problem that appears in some
research task that is important enough to merit investing time to
develop new code for more speed, stability, robustness, or reusability.
One has therefore moved from pathfinding to the construction of a
well-lit paved road, or even a highway (\ref{fig:Production-grade-code}).

\subsection{Intermittent Development \label{subsec:intermit}}

What can we say about the development model and maintainability of
scientific software? This takes us to our next complication: the intermittency
of funding. 

For instance, most electronic structure programs---including both
free and open source packages,\citep{Lehtola2022_WIRCMS_1610} academic
packages, as well as commercial packages---are developed to a significant
extent with public research funding. The intermittency of this funding
causes challenges to sustainability. 

As principal investigators (PIs) usually do not have the resources
(time and/or programming knowledge) to develop or maintain software,
a code development project typically begins with the arrival of a
new student or postdoc. Once the student graduates or the postdoc
leaves for their next position, the code they developed is left without
a maintainer. 

This sustainability problem is further compounded by challenges arising
from the academic environment. Students often have little previous
background in programming---not to mention software engineering---and
usually also only have a superficial understanding of the project.
These two features tend to lead to code that is hard to understand
and hard to maintain. Moreover, sometimes the original author leaves
before the project is finished, and a new student or postdoc takes
their place: rinse and repeat.

Once the initial implementation is finished, in the best case it is
included in a program package. The issue here is that this initial
implementation is rarely written for clarity and for ease of maintenance,
because the first implementation is rarely perfect. Unless significant
development time is later spent to refactor and clean up the initial
implementation---which typically only happens when there is a direct
research based need---this research-level code prevails in the program
package, aging in peace.

In the worst case, the code is not included in any package, which
typically means that the code is simply lost, and the next person
needing this functionality has to start from scratch. This can be
a big issue for the reproducibility of science: as has been argued
by \citet{Hinsen2014_F_101} and further discussed by \citet{Lehtola2022_WIRCMS_1610},
typical software contains hidden aspects that are impossible to describe
in the traditional journal article format; instead, access to the
program source code is often necessary to understand the whole of
the algorithm. The ``war on supercooled water''\citep{Smart2018_PT_}
discussed by \citet{Lehtola2022_WIRCMS_1610} is a good example of
this, as is our recent analysis on the (lack of) reproducibility of
density functional approximations.\citep{Lehtola2023_JCP_114116}

Note that even experienced programmers may have challenges to fulfill
the criteria of modern-day software engineering in the case of research
code, due to the nebulous nature of the task: it is usually not clear
at the outset what the best, simplest and thereby most maintainable
implementation looks like. As new methods are invented and implemented,
design choices that originally looked reasonable may turn out to be
problematic. Research code therefore has a tendency of ``spaghettifying'':
what started out originally as a clean and logical implementation
can turn into a jumbled mess, when suffiently many piecewise changes
are made to the code, especially if the project does not pay a lot
of attention to code quality through coding standards, developer training,
and code review. 

Production grade code, in turn, is usually only achievable when the
author(s) are expert programmers who are also inherently familiar
with the task they are trying to accomplish. Acquiring this experience
typically requires experience with previous implementations of the
same task; as was discussed above, more often than not, the first
try is not pretty, and later refactors are necessary.

As a result of the above features of scientific software development,
scientific software packages---especially ones that are successful---tend
to be formed of collections of layers of code from a variety of authors,
exhibiting large variation in the quality of the code included. 

\subsection{Scoped Development Efforts \label{subsec:scope}}

Let us next discuss how the focus on new developments is typically
assigned. Given that developer time is a scarce resource, the smartest
approach is to focus the efforts where they are most impactful. This
either means focusing on implementing new features in the code, or
to speed up bottlenecks in old code. Importantly, the Pareto principle
also applies in scientific computing. Often, 80\% of the runtime is
spent in 20\% of the code; depending on the context, the ratios can
be even more extreme, such as 95\% of the time being spent on 2\%
of the code. This division is invariably used to allocate development
efforts in scientific programs: focus is given to the performance
critical parts of the code, while the rest of the code is let to age
in peace.

As new algorithms are continuously discovered and published, improved
algorithms make their way into programs as new functionalities are
implemented or old code is rewritten. However, this progress can be
extremely slow. It is essential to note that programs and libraries
reflect their developers' primary interests: typically, the code for
doing the types of calculations that the developers are primarily
interested in is in much better shape than code that is less related
to such tasks.

Since keeping up to date with literature is time-consuming, many low-hanging
fruits may be missed, as developers are not aware of the possibilities
inherent in newer approaches. For example, in the domain of numerical
integration a.k.a. quadrature for density functional theory, it is
evident that many programs employ quadrature rules developed several
decades ago, while more recent rules afford better converged results
at the same cost. Likely explanations for the use of stagnated approaches
include that the original implementation was written by a graduate
student or postdoc who left (\ref{subsec:intermit}) or that the original
developer's interests have moved on. After all, if the old implementation
works, there is little incentive to go through the extra effort to
read and try to understand the old code in order to extend it to the
newer quadrature rules---especially if the developers are not up-to-date
on the best practices in the domain of density functional quadrature.
There is always more work to do than time in academia!

Another example is one on initial guesses for electronic structure
calculations. The superposition of atomic densities\citep{Almloef1982_JCC_385,VanLenthe2006_JCC_32}
(SAD) and the superposition of atomic potentials\citep{Lehtola2019_JCTC_1593,Lehtola2020_JCP_144105}
(SAP) are high-quality guesses that often predict energy level orderings
to at least a qualitative level of precision, which is extremely important
for the efficiency and stability of the solution to the self-consistent
field problem.\citep{Lehtola2020_M_1218} 

Despite the huge importance of the initial guess, many electronic
structure programs still lack access to these modern high-quality
guesses, some even exclusively relying on the use of the core guess
obtained by the diagonalization of the one-electron part of the Hamiltonian.
While simple and easy to implement, the core guess completely disregards
electronic interactions and thereby does not reproduce the shell structure
of atoms, and is accordingly horribly inaccurate.\citep{Lehtola2019_JCTC_1593} 

Although SAD is perhaps the most accurate initial guess, implementing
it correctly in a self-consistent manner is challenging.\citep{Lehtola2020_PRA_12516}
The SAP guess, in turn, is extremely easy to implement by quadrature;\citep{Lehtola2019_JCTC_1593}
it can also be efficiently implemented in Gaussian-basis codes in
terms of a small number of three-center two-electron integrals.\citep{Lehtola2020_JCP_144105}
SAP reproduces atomic shell structures and is orders of magnitude
more accurate than the core guess still used in many programs.\citep{Lehtola2019_JCTC_1593,Lehtola2020_JCP_144105}

Although these are just simple examples, they are depictive of the
status quo and it is more than likely that many more aspects of widely
used programs have similar deficiencies. Simply by adopting the current
best practices throughout various programs, calculations could be
made easier and faster to run in terms of both human and computer
time. However, the issue is that these best practices are at present
hard to propagate between programs, as everything needs to be reimplemented
from scratch in every program.

\subsection{Lack of Interoperability \label{subsec:interop}}

Because most people only use or develop a single package, this has
lead to the wide-ranging issues with the lack of interoperability
between program packages: it is in general not possible to run calculation
A with program X, and use the resulting data to carry out calculation
B in program Y. For example, while some quantum chemistry codes contain
sophisticated \emph{ab initio} methods, they may only have simplistic
self-consistent field algorithms for performing the required Hartree--Fock
calculation to obtain the reference state, combined with the aforementioned
issues with poor initial guesses. This can mean that the hardest part
of running a sophisticated calculation is merely making the initial
Hartree--Fock calculation converge.

Other codes, in turn, may have put much more effort into the Hartree--Fock
solver, allowing rapid and stable determination of the ground-state
orbitals for challenging systems. For example, Psi4\citep{Smith2020_JCP_184108}
and PySCF\citep{Sun2020_JCP_24109} come with a variety of self-consistent
field algorithms and initial guesses. However, these programs do not
have every possible \emph{ab initio} method reported in the literature
to carry out the second step of the calculation. 

Although passing the Hartree--Fock orbitals between programs is sometimes
possible through external software such as MOKIT\bibnote{Jingxiang Zou, Molecular Orbital Kit (MOKIT), https://gitlab.com/jxzou/mokit (accessed 26 August 2023)}
and IOData,\citep{Verstraelen2020_JCC_458} there are several technical
barriers that complicate the interfacing. 

First and foremost, there is no generally agreed single standard for
storing basis sets and wave functions on disk: this is done differently
in every program. While some programs have long featured checkpointing
capabilities (the state of the calculation is saved into a single
file which can be used to restart calculations), many others still
lack such capability. 

Interfacing programs traditionally relies on the use of proprietary
formats, such as the formatted checkpoint files of the GAUSSIAN program,\citep{Frisch2016__}
or the format of the MOLDEN visualization program.\citep{Schaftenaar2017_JCMD_789}
Only recently has there been renewed community efforts into program
agnostic interfaces, such as MDI,\citep{Barnes2021_CPC_107688} QCSchema\citep{Smith2020_WCMS_1491}
and TrexIO.\citep{Posenitskiy2023_JCP_174801} Still, as far as I
am aware, QCSchema does not at present ensure that data such as the
basis set and the molecular orbitals can be passed between programs. 

An important thing to note here are the various issues related to
form and data incompatibilities in quantum chemistry. The former issue
is that different programs have made different choices regarding the
ordering, phase, and normalization of either the Cartesian or the
spherical harmonic\citep{Schlegel1995_IJQC_83} Gaussian basis functions.
Although the contents of the data is the same---for instance, the
molecular orbital expansion coefficients are matrices in all programs---the
interpretation of these data is different between programs, and requires
translating between the various conventions employed by the programs.
The coefficients may need to be reordered and multiplied with phase
and/or normalization factors to work in another program.

The latter issue with Gaussian basis set calculations is the significant
difference between segmented and generally contracted\citep{Raffenetti1973_JCP_4452}
basis sets. Many Gaussian-basis programs are optimized for handling
segmented contractions, while the optimal handling of generalized
contractions requires a wholly different implementation deep down
in the quantum chemistry program. This difference means that even
if data could be freely passed between various programs, the interfacing
might not be worthwhile due to the inefficiencies arising from dissimilarities
in the optimal algorithms for the two use cases.

In summary, there are unfortunately few common standards at the moment.
It is also worthwhile noting that many commercial packages disincentivize
interoperability and collaboration in order to protect commercial
interests.\citep{Jacob2016_JPCL_351} Is there any way forward? Although
many of the following observations will also apply to research software,
as well, the primary focus of the upcoming discussion must be limited
to production grade codes due to the indefinite and constantly evolving
nature of research software. 

\section{Reusable Libraries \label{sec:Reusable-Libraries}}

A common problem behind the issues discussed in \ref{sec:Code-Development}
is that there are too few reusable libraries. First, a production
grade reusable open source library---open source being understood
here within the definition of the Open Source Initiative\citep{OSI__}
(which fits within the definition of free and open source software
used in this work\citep{Lehtola2022_WIRCMS_1610})---can evade the
issue of intermittent development. 

Deprecating code in a even a few program packages with a reusable
library frees up significant developer resources. If only a small
fraction of the freed up resources are spent on improving and maintaining
the reusable library, this allows keeping the library well up to date.

Reusable open source libraries can also attract developers from the
larger community. Indeed, a reusable library employed by several codes
is an extremely attractive target for new methods developers: implementing
your new method in a reusable library will instantly make your improvements
available to a large number of downstream users, while implementing
the method in a single program only helps the users of that program.

The larger pool of developers in a reusable open source library likewise
avoids the issues of intermittent development. If state-of-the-art
open source code development practices such as code review are employed,
this also supports the training of new developers and maintainers
for the project.\citep{Lehtola2022_WIRCMS_1610} 

The issue with scoped development efforts is likewise less of an issue
in production grade open source reusable libraries than similar libraries
embedded within a monolithic package. The former project has a clear
scope, guiding and simplifying its development and maintenance, and
also attracting the attention of world wide subject matter experts
who are inherently familiar with the state-of-the-art in that field.
The clear focus of the project helps to keep it on track. 

Reusable libraries can also be used to address many of the present
issues with interoperability: a rich ecosystem of reusable libraries
can even be used to circumvent this issue altogether. Many use cases
at present arise from being unable to run a complicated calculation
with a single program due to technical challenges. Now, if the necessary
subtasks can be carried out by reusable libraries, one only needs
to link the libraries together in a single program, avoiding the need
to interface different programs. 

\subsection{Difference to Traditional Programs \label{subsec:Difference-to-Traditional}}

It is essential to remark here on the difference between modular and
reusable software. Because distributing software used to be difficult---requiring,
for example, sending stacks of floppies in the mail---it was also
difficult to synchronize development efforts in various groups around
the world. This naturally lead to the ``silo'' model,\citep{Oliveira2020_JCP_24117}
in which everything is done in-house, in the program. Unfortunately,
many program packages are still developed within this approach; yet,
they are all already modular. Modular design (functionalities are
implemented with the help of independent modules) has been one of
the leading principles of software design for an extremely long time,
and it has also been long discussed in the computational chemistry
literature: for example, already in \citeyearpar{Bunge1988_CC_85}
\citet{Bunge1988_CC_85} discussed modular libraries as a solution
of the software crisis hindering innovation in electronic structure
calculations---and yet this crisis still persists, as is evident
by the large amount of duplicated effort and obsolete code across
program packages. 

The problem here is that the main focus of modular design is typically
on the silo itself---such as the electronic structure program developed.
Modules in ``siloed'' packages are often little more than glorified
compilation units: they are used to enhance the maintainability of
the silo, while interoperability is not a consideration at all. An
overwhelming majority of recently introduced program packages are
still developed within this silo mentality.

Although all reusable software is modular, not all modular software
is reusable. Modularity and interoperability---the theme of this
Journal of Chemical Physics special issue---are merely two aspects
of reusable software, which is what we are missing. The key thing
to recognize is that there are many tasks in electronic structure
calculations, for example, where the algorithm is generally applicable
and could be implemented in a reusable library.

\subsection{Benefits \label{subsec:Benefits}}

Having an ecosystem of reusable libraries for common tasks accelerates
science by allowing more rapid development of new programs and methods
by new combinations of pre-existing libraries, and significantly reduces
the maintenance overhead.\citep{Tracz1988_ASSEN_17,Poulin1993_ISJ_567}
Instead of having to maintain several dozens of independent implementations
in as many packages, many of which were not developed in a maintainable
fashion (\ref{subsec:intermit}), an ecosystem relying on shared reusable
open-source libraries allows the elimination of redundant code across
packages and thereby requires less developer effort in total. As open
source software is infinitely replicatable, it has been recognized
in the economic literature as a public good:\citep{Myatt2002_OREP_446,Johnson2002_JEMS_637,Krogh2006_MS_975}
it is a service to the public that does not become scarcer when used.

Reusable libraries are described by the four following characteristics:
(i) separation of concerns, (ii) high cohesion, (iii) loose coupling,
and (iv) information hiding. Separation of concerns implies modularity:
a given module only does a given thing. The three further criteria
go beyond mere modularity. High cohesion means that all elements in
the library strongly belong together. Loose coupling means that the
library has few external dependencies, while information hiding means
that the developers of the library have free rein to improve the implementation
behind the scene, for instance by adopting a more efficient algorithm.

Unlike some decades ago, reusable open source libraries are nowadays
a real option, as floppy disks in the mail have been replaced by distributed
version control systems such as \emph{git} which efficiently enable
independent collaboration of people around the world. In fact, \emph{git}
was developed to enable the decentralized development of the Linux
operating system kernel by tens of thousands of developers, and it
appears to already be used by a majority of electronic structure packages.

\subsection{Sustainability Crisis and Opposition \label{subsec:Sustainability-Crisis-and}}

It is, however, hard to break out of the silo model of thinking. My
conviction for reusable libraries has arisen from decades of work
across different program packages. I wrote the Gaussian-basis ERKALE
program\citep{Lehtola2012_JCC_1572} during my PhD, and the finite
element method HELFEM program\citep{Lehtola2019_IJQC_25945,Lehtola2019_IJQC_25944,Lehtola2020_MP_1597989,Lehtola2020_PRA_12516,Lehtola2023_JCTC_2502,Lehtola2023_JPCA_4180}
at the start of my independent career. In addition, I have made contributions
to Q-Chem,\citep{Epifanovsky2021_JCP_84801} Psi4,\citep{Smith2020_JCP_184108}
PySCF,\citep{Sun2020_JCP_24109} and OpenMOLCAS.\citep{Manni2023_JCTC_}
Already many years ago, I awoke to a sustainability crisis: I realized
I could not implement my methods in all of these programs, as it would
not simply be worth the required time investment. Moreover, maintenance
of the duplicate implementations would become an absolute nightmare.

Excluding the proprietary Q-Chem package, the remaining five free
and open source packages still contain large amounts of duplicated
code that could be stripped apart into reusable libraries. I see little
benefit in solving individual issues in individual codes, when instead
it should be perfectly possible to use a reusable open source library
in all of these programs. This problem is made ever so more acute
by my research plan, which focuses on the development of new numerical
atomic orbital methods.\citep{Lehtola2019_IJQC_25968} Traditionally,
this would require writing yet another program package from scratch,
which would further complicate the maintainability crisis.

A typical response to my complaint on the lack of reusable implementations
is that I could already accomplish my research project simply by carrying
out my project in a given modular program package like PySCF. Although
well-intended, such a simple suggestion omits quintessential drawbacks. 

Such an interface would be heavily suboptimal. While it would suffice
for toy calculations, a good proof-of-concept study for a novel approach
requires demonstrating that it scales to large systems, which requires
end-to-end considerations.

Such an implementation within a monolithic (while modular) package
would also be limited by the issue of the lack of interoperability.
It is both in my interest as the developer and in that of the greater
community that any new methods I develop become accessible to the
largest number of people. An implementation tied to a given program
package will only merit the users of that package, availing the method
to users of other packages again requiring duplicated efforts in these
other packages. 

Instead, a reusable open source implementation solves the problem
once and for all, promoting easy adoption in all programs. This is
my main motivation and rationale for pursuing new reusable libraries.
Although I could keep working in the ways of old, I believe that the
need to develop a new code also allows changing the modus operandi
with little extra effort. I have also received enthusiastic feedback
for this proposal from colleagues in industry: there is a significant
unsatisfied need for reusable libraries in quantum chemistry. 

\subsection{Example: Libxc \label{subsec:Example:-Libxc}}

What do projects in the envisioned ecosystem look like? Our Libxc
library of density functional approximations\citep{Lehtola2018_S_1}
(DFAs) is the epitome of a successful reusable library in electronic
structure theory. Now used by roughly 40 electronic structure packages,
including both all-electron, pseudopotential, and projector augmented
wave approaches based on various numerical representations such as
atomic orbitals, plane waves, finite elements, finite differences,
and multiresolution grids, Libxc has become the standard implementation
of density functionals, greatly enhancing the reproducibility of electronic
structure calculations.\citep{Lehtola2023_JCP_114116} 

The key to Libxc's success is that it actually encapsulates all of
the four above traits of reusable software. Libxc only handles density
functional approximations. It evaluates the density functional $f_{\text{xc}}$
in the semi-local approximation of the exchange-correlation (xc) energy
\begin{equation}
E_{\text{xc}}=\int f_{\text{xc}}(n,\nabla n,\dots){\rm d}^{3}r\label{eq:Exc}
\end{equation}
as well as the derivatives of $f_{\text{xc}}$ needed to compute derivatives
of $E_{\text{xc}}$ (separation of concerns and high cohesion; kinetic
energy approximations for orbital free DFT are also supported by Libxc).
In the same spirit, Libxc also supplies the other data needed to evaluate
the density functional approximation, such as the mixing coefficients
for exact exchange for hybrid functionals, and range-separation parameters
for range-separated hybrids. To continue, only a minimal amount of
information needs to be passed between Libxc and the calling program
(loose coupling), and the routines that are used to calculate the
values of $f_{\text{xc}}$ and its derivatives can be freely modified
(information hiding). 

It needs to be noted that Libxc was not the first modular implementation
of density functionals: before the first publication on Libxc by \citet{Marques2012_CPC_2272},
libraries of DFAs had been described by \citet{Strange2001_CPC_310},
\citet{Salek2007_JCC_2569}, and \citet{Ekstroem2010_JCTC_1971},
for example. With the exception of the XCFun library of \citet{Ekstroem2010_JCTC_1971},
it appears that none of these projects gained wide adoption for reasons
unknown to the present author.

Libxc originated in the Octopus community\citep{TancogneDejean2020_JCP_124119}
which is interested in finite differences calculations with pseudopotentials,
but already in \citeyearpar{Marques2012_CPC_2272} it was used by
13 programs.\citep{Marques2012_CPC_2272} Although many fewer codes
used Libxc 10 years ago than now, it is clear that the user community
was already sufficiently diverse to result in various improvements
to the code, which later resulted in its present near-ubiquitous status.
Especially, ERKALE\citep{Lehtola2012_JCC_1572,Lehtola2018__a} was
one of the first Gaussian-basis codes to interface Libxc, and the
first program to fully use its hybrid functional interface, motivating
this author's initial contributions to Libxc. Interestingly, Gaussian-basis
codes have now become the most numerous users of Libxc.

\section{Promises and Challenges \label{sec:Promises-and-Challenges}}

If a given task is standardized behind an application programming
interface (API), there is great benefit to both developers and end
users. An extreme example is the case of the basic linear algebra
subsystem (BLAS) discussed by \citet{Lehtola2022_WIRCMS_1610}. It
is noteworthy that while BLAS is nowadays ubiquitous,\citep{Perkel2021_N_344}
it took years of concerted efforts by the BLAS team to convince both
academia and industry to adopt the standard. 

The interoperability of various implementations of the same API allowed
experts to focus on ways to improve the specific tasks in BLAS, and
made these implementations available to all end users: a large number
of interoperable BLAS libraries are available, ranging from the Netlib\bibnote{https://www.netlib.org/, Accessed 16 October 2023.}
reference implementation to academic projects like the BLAS-like Library
Instantiation Software Framework\citep{Zee2015_ATMS_1} (BLIS) to
vendor optimized libraries like Intel's Math Kernel Library (MKL),
AMD's Optimizing CPU Libraries (AOCL), and NVidia's Compute Unified
Device Architecture (CUDA) math library. 

The benefit of standard APIs also extend to scientific software. In
a virtuous cycle, the specification of a shared API allows both subject
matter experts to focus their efforts into improved implementations
of that API, and makes these improvements available to various program
packages. A standard implementation is attractive to supercomputing
centers and vendors, who have an easier time to optimize a single
library than optimizing dissimilar implementations in a multitude
of programs. This also simplifies targetting developing computer hardware.
Computational bottlenecks are often similar across numerical algorithms,
enabling the leveraging of reusable implementations to solve performance
and scaling issues for the same task in a multitude of codes. Shared
open source implementations just are more maintainable.

In this aspect, it is interesting to compare traditional scientific
software to machine learning software. All the major machine learning
packages, such as TensorFlow,\citep{MartinAbadi2015__} Keras,\citep{Chollet2015__}
scikit-learn,\citep{Pedregosa2011_JMLR_2825} OpenNN,\citep{Martin2021__}
Theano,\citep{TDT2016__} and PyTorch\citep{Paszke2019__} are open
source software. In machine learning, there is little competitive
advantage in the implementation of the machine learning model; instead,
it is the data that is used to train the models that is valuable for
commercial companies. It is then evident that having to maintain and
develop the software is just unnecessary overhead.

Free and open source software has been gaining commercial adoption
also in computational chemistry. For example, the Open Force Field
initiative is an industrially funded consortium for open science that
targets the development of new open source force fields, including
also the development of the open source software needed to (i) fit
such force fields and (ii) to use them in calculations.\citep{Qiu2021_JCTC_6262,Horton2022_JCIM_5622,Boothroyd2022_JCTC_3566,Boothroyd2023_JCTC_3251}
Access to pre-existing free and open source packages such as Psi4\citep{Smith2020_JCP_184108}
is key to this effort, as it enables building of new methods and tools
on top of existing infrastructures.

An often-encountered misunderstanding with open sourcing your software
is that in doing so, you are giving your work to your competitors
for free. However, the reality is often the complete opposite: open
sourcing can mean extracting free development work from the community.
There is some grain of truth in the software companies' beloved statement
``free and open source software is only free if you don't value your
time''.

Making small bug fixes often takes significant developer effort from
project maintainers, simply since a package may have many small bugs
that only occur in rare instances. In a proprietary package, all of
this effort lands on the shoulders of the full-time developers, which
means that addressing all the small bugs can take considerable developer
resources. In contrast, in an open source project, any end user experiencing
a bug may end up sufficiently motivated to fix the bug themselves.
Indeed, contributing such bug fixes is often the way how open source
projects gain new contributors (as is also true about this author's
original involvement with Libxc,\citep{Lehtola2018_S_1} Psi4,\citep{Smith2020_JCP_184108}
and PySCF,\citep{Sun2020_JCP_24109} for instance).

What many fail to note is that such transient contributions from the
community can together account to a significant amount of developer
time, which obviates the need to have developers on the payroll to
work on such small maintenance tasks. This again takes us back to
the public good aspect of free and open source software: improvements
made by private parties are available to everyone.\citep{Johnson2002_JEMS_637} 

This discussion would not be complete without noting also the economic
motivation for individuals to make such contributions, in addition
to their own vested interest (e.g. getting their calculation to run
without hitting a bug). Unlike work done in proprietary packages,
contributions to free and open source packages are visible to everyone,
allowing potential future employers or clients an unfettered view
on the coding and collaboration skills of the individual developer:
many free and open source software developers are recruited for their
talents by commercial software companies.

A key inhibition for the adoption of reusable libraries is the (feeling
of) loss of control that relying on an external library comes with.
Before adopting an external library, potential users typically want
to see that the project is actively maintained: there should be a
constant stream of development, bug fixes, and predictable release
management. Users can also worry about longer term sustainability:
is the library going to be (i) maintained and (ii) developed for the
foreseeable future? Even excellent quality reusable libraries may
be hard to ``sell'' to program packages, if they have been developed
by a single academic who does not have a permanent position.

Actively developed packages may have other hindrances in the world
of commercial program packages, which are especially wary of anything
that could potentially affect their ability to deliver a standard
quality product to their customers. Commercial packages typically
do not like if libraries change their algorithms, even if the new
algorithm is better than the old one, as all minor releases of a given
major release of a program package should reproduce the same results
(barring bugs that affect the result). Too active development can
also be seen as a bad thing!

However, this loss of control can be avoided through careful version
control and the ability to control the operation of the library: pre-existing
critical functionality such as old algorithms should not be instantly
removed, but kept as an option for a considerable migration period.

It must also be stated that the mere existence of a reusable implementation
does not force anyone to use it. Even with the exclusion of commercial
packages, there remain many academic as well as free and open source
projects\citep{Lehtola2022_WIRCMS_1610} that could leverage reusable
libraries in order to radically reduce the time necessary to develop
new techniques, as well as to reduce maintenance effort through deprecation
and elimination of old implementations replaced by a reusable library.
When reusable libraries gain more adoption, they thereby often also
acquire more developers and can enter a virtuous loop of better maintenance
and enhanced adoption. Already the Libxc example shows that one does
not need to get everyone on board at departure; most of the projects
that now use Libxc jumped onboard only relatively recently.

Unfortunately, a common challenge in academia is that reusing others'
software is typically seen as less rewarding to one's career than
writing your own implementation from scratch: there is a rewards system
for reinventing the wheel. This is especially an issue in tasks for
which it is easy to write a simple implementation. 

However, even in such cases, there is still benefit for reusable libraries:
if a reusable open-source library offers more flexibility or variety
in methodology, it is appealing even for simple tasks. For example,
another major reason for the success of Libxc is that while it is
relatively simple to implement the few popular density functionals
like Becke's 1988 exchange (B88) functional,\citep{Becke1988_PRA_3098}
Perdew's 1986 correlation (P86) functional,\citep{Perdew1986_PRB_8822}
the Perdew--Burke--Ernzerhof (PBE) functional,\citep{Perdew1996_PRL_3865,Perdew1997_PRL_1396}
and the Tao--Perdew--Staroverov--Scuseria (TPSS) functional,\citep{Tao2003_PRL_146401,Perdew2004_JCP_6898}
it is much more difficult to implement the 600+ functionals currently
available in Libxc, and to keep adding in more functionals as they
are published.

Unfortunately, the unavoidable compromise between flexibility and
maintainability needs to be taken into account in reusable software.
Even if the issues with interfacing various languages (\ref{sec:Introduction})
may end up requiring separate object-oriented libraries in C++, Fortran,
and Julia, for instance, the reduction from dozens of independent
and redundant implementations to three reusable libraries would still
represent a significant step forward in maintainability, and promote
innovation through the competition of powerful independent implementations.

\section{Summary and Discussion \label{sec:Summary-and-Discussion}}

In \ref{sec:Introduction}, I discussed the aging nature of all software
and the challenges that keeping up with ever-developing theoretical
methods, programming languages, as well as computing hardware pose
on scientific software developers and maintainers. \Cref{sec:Code-Development}
was dedicated to the discussion of the challenges that the academic
environment causes on the maintainability of software. Research software
is by definition unmaintainable, since the goals keep on changing,
and research software is therefore in a constant state of flux. The
focus of any discussion on reusable software must be on production
grade software. Still, the status quo is that dozens of packages attempt
to maintain duplicated codes.

These duplicated codes could be easily replaced with production grade
reusable libraries, whose benefits I discussed in \ref{sec:Reusable-Libraries}.
The elimination of the duplicated effort across packages would address
sustainability issues related to the intermittent nature of academic
code development and funding, and allow domain experts to maintain
a single or a few competing reusable libraries solving dedicated tasks.
Such reusable libraries would also lessen the impact of the long-standing
issues with lack of interoperability between program packages, as
the same sets of functionalities could be imported in all packages
from the same sets of reusable open source libraries. 

I also highlighted the differences of the software development model
based on reusable software to the status quo in many electronic structure
packages: a monolithic package built of modules is still a monolithic
package. Reusability goes well beyond modularity, and reusable software
does not appear to have been given adequate attention in the computational
chemistry community so far. As our Libxc library\citep{Lehtola2018_S_1}
demonstrates, reusable open source libraries can be extremely successful.

The success of the pioneering innovation made with Libxc has already
been followed by the CECAM {[}Centre Européen de Calcul Atomique et
Moléculaire---European center for atomic and molecular calculations{]}
Electronic Structure Library (ESL).\citep{Oliveira2020_JCP_24117}
Although code reuse is mentioned two times in \citeref{Oliveira2020_JCP_24117},
its main focus is still the modular software development paradigm.
The main argument of this Perspective that one should focus on reusability,
not mere modularity: Libxc satisfies all the four properties of reusable
software (\ref{sec:Reusable-Libraries}), which suggests that one
should start the move to reusable open source libraries by the identification
of similar tasks that fit the same criteria. 

For instance, an important consideration for reusable libraries is
that following the characteristics of \ref{sec:Reusable-Libraries},
reusable libraries should be loosely coupled. This means that they
should have minimal dependencies, which is greatly beneficial for
their potential inclusion in program packages. In contrast, software
that is merely modular can be tightly coupled with many dependencies,
leading to complicated dependency trees, software that is harder to
build, and a software stack that breaks more easily.

Finally, in \ref{sec:Promises-and-Challenges}, I discussed some promises
and challenges of reusable software. To continue, perhaps the hardest
challenge for reusable libraries is that their development and maintenance
is not adequately recognized in academia: it can be difficult to accrue
funding for reusable libraries. Also, even though reusable libraries
can be critical in enabling new science, this is not always reflected
in the literature. For instance, even though Libxc is used by at least
40 program packages, many of which accrue thousands to tens of thousands
citations a year, the \citeyearpar{Lehtola2018_S_1} publication on
Libxc\citep{Lehtola2018_S_1} has only been cited a total of 423 times
(Google Scholar, 16 October 2023) in five years. Given that most electronic
structure calculations are carried out with density functionals, which
are almost exclusively supplied by Libxc in most programs, were the
library properly cited in each publication employing it, the library
would likely accrue thousands of citations each year. It is also interesting
to note in this context that the number of citations also pales in
respect to the number of density functionals (>600) available in the
current version of Libxc.

Despite all of these issues, it is my firm belief that reusable libraries
have a bright future in computational chemistry. There are a variety
of tasks where reusable libraries can be fashioned and leveraged to
deprecate obsoleted code in various packages. Reusable libraries can
enhance the reproducibility of calculations across programs, for instance
by standardizing technical aspects such as quadrature grids, which
are currently custom in each program package, which has lead to issues
in the reproduction of calculations; work in this direction is already
ongoing. 

It is for the above reasons that I present my call to arms to the
community: we need more reusable code instead of more modular code.
Although reusability is a harder nut to crack than modularity---a
good scope for a reusable library should satisfy all four criteria
discussed in \ref{sec:Reusable-Libraries}, and the library should
be thoroughly documented and tested---the rewards will likely also
be greater thanks to the greater applicability and impact of such
a project.

Even though much of the discussion above has related to scientific
software, similar observations apply also to industry, where many
applications likewise require code development. Open source software
is successfully employed also in industry. Our Libxc library (\ref{subsec:Example:-Libxc})
is anecdotal proof of this: in addition to various academic as well
as free and open source programs, Libxc is also used by a number of
commercial electronic structure packages. I wholeheartedly recommend
the use of permissive open source licenses, such as the one used in
Libxc, in order to tackle the maintainability crisis in computational
chemistry software. To avoid reinventing the wheel, reusable implementations
have to be usable also by commercial software companies: after all,
the more reusable the code is, the greater its impact will be.

Finally, although the main focus of the perspective was on production
grade software, I would like to reiterate that many of the observations
made in this work also apply to research codes. Most importantly,
a thriving ecosystem of reusable codes also accelerates the development
of research codes: instead of having to reinvent the wheel, to continue
the analogy, one can focus on developing lighter and more durable
spokes, inner tyres that can withstand larger pressures, and better
outer tyre tread patterns for increased fuel efficiency, for instance.
A thriving ecosystem of reusable software enables faster development
cycles, as development can be focused on only the interesting bits
of a given computational problem.

While the issues with sustainable software development discussed in
this work have been well-known for a very long time,\citep{Bunge1988_CC_85}
I believe that the time is finally ripe for a paradigm shift. In my
view, the key issue historically holding back an ecosystem based on
reusable libraries in science has been software maintenance. To be
successful, a reusable library has to be well-maintained: per the
discussion in the Introduction, the library needs to be kept up to
date with evolving algorithms, programming models, and hardware support.
Bugs have to be fixed in a reasonable time frame, and the bugfixes
have to be predictably rolled out into stable releases. As distributing
software used to be difficult, this naturally led to monolithic designs,
where even code that could easily be reusable was forked into incompatible
versions in various packages. However, with the recent advent of distributed
version control systems such as \emph{git}, a different modus operandi
is possible.

Instead of a single monolithic repository for a software package,
many projects already use source version control systems that support
submodules that enable seamless co-operation between various projects,
and use pristine upstream source repositories for their software dependencies.
Combined with the ease of decentralized development with \emph{git},
it is now perfectly feasible for various projects around the world
to collaborate on using and developing a shared set of core libraries
for core tasks. 

In my view, the change has already begun to take place: there already
exist a number of libraries\citep{Kaliman2013_JCC_2284,Wouters2014_CPC_1501,Sun2015_JCC_1664,Remigio2019_IJQC_25685,Scheurer2019_JCTC_6154,Herbst2020_WIRCMS_1462,Shaw2021_JOSS_3039,Asadchev2023_JCTC_1698,WilliamsYoung2023_JCP_214109,Delcey2023_WCMS_1675}
which have been designed to be (re)used in several electronic structure
packages. What we need is more of the same. I therefore urge developers
across program packages to identify shared problem areas suitable
for novel reusable solutions. While focusing on the development of
a reusable library may be a net negative for a single individual or
research group, requiring more investment in software development
than hacking together a typical one-off implementation in the traditional
development model, previous experiences show that a compact number
of developers interested in the same topic but working in different
electronic structure programs can fashion unified state-of-the-art
solutions which are highly reusable.

Indeed, I believe that the lowest hanging fruit and the best potential
solution to the present problems to lie in novel collaborations between
various developer communities that have so far not been in strong
interaction. It only takes a small group of people collaborating across
program packages to reach a critical mass to sustain a reusable library,
which can later snowball into a \emph{de facto} industry standard.
I hope to follow up this work with reports of various reusable libraries
for electronic structure theory, the first of which should appear
in this special issue.
\begin{acknowledgments}
This work is heavily motivated by the work I did and the discussions
I had while working at the Molecular Sciences Software Institute at
Virginia Tech from 2020 to 2022. Out of the many people I have spoken
with and had enlightening discussions on scientific software development,
I especially want to thank Paul Saxe, David Williams-Young, Edward
Valeev, Martin Head-Gordon, Frank Neese, Peter Knowles, Jeff Hammond,
and Christopher Bayly. I also thank David and Jeff for discussions
and feedback on this manuscript. I also thank the Academy of Finland
for financial support under project numbers 350282 and 353749.
\end{acknowledgments}

\bibliography{citations}

\end{document}